\documentstyle[prd,aps]{revtex}

\newcommand{\bea}{\begin{eqnarray}}
\newcommand{\eea}{\end{eqnarray}}

\begin{document}

\draft
\twocolumn[\hsize\textwidth\columnwidth\hsize\csname
@twocolumnfalse\endcsname
\title{Quantum fluctuations of Cosmological Perturbations 
       in Generalized Gravity}
\author{Jai-chan Hwang}
\address{Department of Astronomy and Atmospheric Sciences,
         Kyungpook National University, Taegu, Korea}
\date{\today}
\maketitle

\begin{abstract}

Recently, we presented a unified way of analysing classical 
cosmological perturbation in generalized gravity theories.
In this paper, we derive the perturbation spectrums generated 
from quantum fluctuations again in unified forms.
We consider a situation where an accelerated expansion phase of 
the early universe is realized in a particular generic phase of 
the generalized gravity.
We take the perturbative semiclassical approximation which treats 
the perturbed parts of the metric and matter fields as quantum 
mechanical operators.
Our generic results include the conventional power-law and exponential 
inflations in Einstein's gravity as special cases.

\end{abstract}

\noindent
\pacs{PACS numbers: 03.65.Sq, 04.50.+h, 04.62.+v, 98.80.Cq}

\vskip2pc]
The highly isotropic cosmic microwave background radiation permits 
a linear paradigm of treating the structure formation process in the 
evolving universe.
It is widely accepted that the currently observable large scale 
structures are developed from remnants of the quantum fluctuations 
imprinted during early accelerated expansion stage.
In the context of Einstein's gravity both the quantum generation and 
classical evolution processes can be rigorously handled; 
studies in \cite{H-QFT,H-IF} can be compared with the previous order 
of magnitude analyses \cite{pert-infl-rough}, or analyses in differnt 
gauges \cite{pert-infl-other}.
For a scalar field, self consistent and rigorous analyses are possible
mainly due to a special role of particular gauge condition 
(or equivalently, gauge invariant combination) which suits the problem:
the uniform-curvature gauge.

Variety of theoretical reasons allude possible generalization of 
the gravity sector due to quantum correction in the high energy limit, 
and thus in the early universe \cite{Quantum-correction,QFCS}.
There were many studies on the classical evolution of structures
in some favored generalized gravity theories \cite{pert-GGT-infl}.
However, again, in the classes of generalized gravity involving the scalar 
field and scalar curvature, we found that the uniform-curvature gauge suits
the problem allowing simple and unified treatment possible \cite{GGT-H}.
For the growing mode, the large scale solution known in the minimally 
coupled scalar field remains valid even in a wide class of generalized 
gravity.
In this paper we investigate the quantum generation process in
the context of generalized gravity.
The properly chosen gauge condition again allows us to present the 
generated power spectrum in generic forms which are applicable to various 
generalized gravity theories and underlying background evolutions.

We consider the gravity theories with an action
\bea
   & & S = \int d^4 x \sqrt{-g} \left[ {1\over 2} f(\phi, R)
       - {1 \over 2} \omega(\phi) \phi^{;a} \phi_{,a} - V(\phi) \right],
   \label{Action-GGT}
\eea
where $\phi$ is a scalar field and $R$ is a scalar curvature;
$f$ is a general algebraic function of $\phi$ and $R$,
and $V$ and $\omega$ are general functions of $\phi$.
It includes diverse classes of generalized gravity theories 
as subsets, \cite{GGT-HN,GGT-CT}.

As a metric describing the universe we consider a spatially homogeneous, 
isotropic and flat background, and general perturbations of a scalar type
\bea
   d s^2 
   &=& - \left( 1 + 2 \alpha \right) d t^2 - \chi_{,\alpha} d t d x^\alpha
   \nonumber \\
   & & + \; 
       a^2 \delta_{\alpha\beta} \left( 1 + 2 \varphi \right)
       d x^\alpha d x^\beta,
   \label{metric}
\eea
where $a(t)$ is a cosmic scale factor; $\alpha ({\bf x}, t)$, 
$\chi ({\bf x}, t)$, and $\varphi ({\bf x}, t)$ are perturbed order 
quantities.
{}For the scalar field we let
\bea
   & & \phi ({\bf x}, t) = \bar \phi (t) + \delta \phi ({\bf x}, t),
\eea
where the background quantities are indicated by overbars 
which will be neglected unless necessary.
We have not chosen the temporal gauge condition which can be used
as an advantage in handling problems; all perturbed order variables
are spatially gauge invariant, see Sec. IV C of \cite{GGT-HN}.
We introduce a gauge invariant combination
\bea
   & & \delta \phi_\varphi
       \equiv \delta \phi - {\dot \phi \over H} \varphi
       \equiv - {\dot \phi \over H} \varphi_{\delta \phi},
   \label{UCG-UFG}
\eea
where $H \equiv \dot a / a$.
$\delta \phi_\varphi$ is a gauge invariant variable which is the same as 
$\delta \phi$ in the uniform-curvature gauge which chooses $\varphi =0$
as the gauge condition; Eq. (\ref{metric}) shows that $\varphi = 0$
implies that spatial curvature vanishes (uniform in general),
thus justifying the name. 

Ignoring the surface terms, the action valid to the second order in the 
perturbation variables is derived in \cite{GGT-CT} as
\bea
   \delta S
   &=& {1\over 2} \int a^3 Z \Bigg\{ \delta \dot \phi_\varphi^2
       - {1 \over a^2} \delta \phi_\varphi^{\;\; |\alpha}
       \delta \phi_{\varphi,\alpha} 
   \nonumber \\
   & & \qquad + \;
       {1 \over a^3 Z} {H \over \dot \phi} 
       \left[ a^3 Z \left( {\dot \phi \over H} \right)^\cdot \right]^\cdot
       \delta \phi_\varphi^2 \Bigg\} dt d^3 x.
   \label{Action-pert}
\eea
The non-Einstein nature of the theory is present in a parameter $Z$
which is defined as
\bea
   & & Z (t) \equiv { \omega + {3 \dot F^2 \over 2 \dot \phi^2 F } 
       \over \left( 1 + {\dot F \over 2 H F} \right)^2 },
\eea
where $F \equiv \partial f/ (\partial R)$.
$Z$ becomes unity in Einstein's gravity where $F = 1 = \omega$.
Equation (\ref{Action-pert}) leads to an equation of motion 
\bea
   \delta \ddot \phi_\varphi 
   &+& { (a^3 Z)^\cdot \over a^3 Z }
       \delta \dot \phi_\varphi 
   \nonumber \\
   &-& \left\{ {1 \over a^2} \nabla^2
       + {1\over a^3 Z} {H \over \dot \phi} \left[ 
       a^3 Z \left( {\dot \phi \over H} \right)^\cdot \right]^\cdot
       \right\} \delta \phi_\varphi = 0.
   \label{GGT-delta-phi-eq} 
\eea
In the {\it large scale limit}, ignoring the Laplacian term, we have
a general integral form solution 
\bea
   & & \delta \phi_\varphi ({\bf x}, t)
       = - {\dot \phi \over H} \left[ C ({\bf x}) - D ({\bf x}) \int_0^t
       {1 \over a^3 Z} {H^2 \over \dot \phi^2} dt \right],
   \label{UCG-delta-phi-LS-sol}
\eea
where $C({\bf x})$ and $D({\bf x})$ are coefficients of the growing
and the decaying modes, respectively.
Notice that the growing mode is not affected by the non-Einstein nature 
of the theory.
The growing mode of $\varphi_{\delta \phi}$ is conserved as $C({\bf x})$,
whereas the decaying mode is higher order in the large scale expansion;
see Sec. VI A of \cite{GGT-HN}.
Thus, $C({\bf x})$, which encodes the spatial structure, can be 
interpreted as $\varphi_{\delta \phi}$ in the large scale limit.
The solution in Eq. (\ref{UCG-delta-phi-LS-sol}) is valid considering 
general $V(\phi)$, $\omega(\phi)$, and $f(\phi,R)$ in various subsets 
of generalized gravity theories described in \cite{GGT-HN,GGT-CT}.

Introducing
\bea
   & & v ({\bf x}, t) \equiv z { H \over \dot \phi } \delta \phi_\varphi, 
       \quad z (t) \equiv {a \dot \phi \over H} \sqrt{Z},
   \label{v-def}
\eea
Eq. (\ref{GGT-delta-phi-eq}) can be written as
\bea
   & & v^{\prime\prime} - \left( \nabla^2
       + {z^{\prime\prime} \over z} \right) v = 0,
   \label{v-eq}
\eea
where a prime denotes a derivative with respect to the conformal time $\eta$,
$d \eta \equiv dt/a$.
In the {\it small scale limit} ($z^{\prime\prime}/z \ll k^2$) we have
\bea
   & & \delta \phi_\varphi ({\bf k}, \eta)
       = { 1 \over a \sqrt{2k} } \Big[ c_1 ({\bf k}) e^{i k \eta}
       + c_2 ({\bf k}) e^{-ik\eta} \Big] {1 \over \sqrt{Z} }.
\eea

In an approach called a {\it perturbative semiclassical approximation},
\cite{H-UCG}, we regard the perturbed part of the field and metric as 
quantum mechanical operators, meanwhile the background parts are 
considered as classical.
Instead of the classical decomposition we replace the perturbed order 
variables with the quantum (Heisenberg representation) operators as
\bea
   & & \phi ({\bf x}, t) = \bar \phi (t) + \delta \hat \phi ({\bf x}, t), 
       \quad
       \varphi ({\bf x}, t) \rightarrow \hat \varphi ({\bf x}, t), \quad 
       {\rm etc.}, 
   \nonumber \\
   & & \delta \hat \phi_\varphi \equiv \delta \hat \phi 
       - {\dot \phi \over H} \hat \varphi.
   \label{decomposition-quantum}
\eea
An overhat indicates the quantum operator.
The background order quantities are considered as classical variables.
This approach has a different spirit compared with the quantum field
theory in curved spacetime, where in the latter case the metric sector
is regarded as classical and given (sometimes considering some prescribed
backreaction) \cite{QFCS,Duff}.
Our approach considers the field and the metric in equal footing 
\cite{H-UCG}.
Since we are considering a {\it flat} three-space background we may expand
$\delta \hat \phi ({\bf x},t)$ in the following mode expansion
\bea
   \delta \hat \phi_\varphi ( {\bf x}, t) 
   &=& \int {d^3 k \over ( 2 \pi)^{3/2} }
       \Big[ \hat a_{\bf k} \delta \phi_{\varphi {\bf k}} (t) 
       e^{i {\bf k}\cdot {\bf x}}
   \nonumber \\
   & & \qquad \qquad \qquad+ \;
       \hat a^\dagger _{\bf k} \delta \phi^*_{\varphi {\bf k}} (t) 
       e^{-i {\bf k} \cdot {\bf x}} \Big].
   \label{mode-expansion}
\eea
The annihilation and creation operators $\hat a_{\bf k}$ and
$\hat a_{\bf k}^\dagger$ satisfy the standard commutation relations:
\bea
   & & [ \hat a_{\bf k} , \hat a_{{\bf k}^\prime} ] = 0, \quad
       [ \hat a^\dagger_{\bf k} , \hat a^\dagger_{{\bf k}^\prime} ] = 0, 
       \quad
       [ \hat a_{\bf k} , \hat a^\dagger_{{\bf k}^\prime} ]
       = \delta^3 ( {\bf k} - {\bf k}^\prime ).
   \nonumber \\
   \label{annihilation}
\eea
$\delta \phi_{\varphi {\bf k}} (t)$ is a mode function, a complex 
solution of the classical mode evolution equation.
Equation (\ref{GGT-delta-phi-eq}), with $\delta \hat \phi_\varphi$ 
replacing $\delta \phi_\varphi$, leads to an equation for the mode 
function $\delta \phi_{\varphi {\bf k}}$ which satisfies the same form as
Eq. (\ref{GGT-delta-phi-eq}).

{}From the action for $\delta \phi_\varphi$ in Eq. (\ref{Action-pert})
we can derive the conjugate momenta and the commutation relation.
Using $S = \int {\cal L} dt d^3 x$ we have
$\delta \pi_\varphi ({\bf x}, t) 
\equiv \partial {\cal L} / (\partial \delta \dot \phi_\varphi) 
= a^3 Z \delta \dot \phi_\varphi ({\bf x}, t)$.
Thus, the equal-time commutation relation
$[\delta \hat \phi_\varphi ({\bf x},t),
\delta \hat \pi_\varphi ({\bf x}^\prime, t) ]
= i \delta^3 ({\bf x} - {\bf x}^\prime)$ leads to
\bea
   & & [ \delta \hat \phi_\varphi ({\bf x},t),
       \delta {\dot {\hat \phi}_\varphi} ({\bf x}^\prime, t) ]
       = {i \over a^3 Z} 
       \delta^3 ({\bf x} - {\bf x}^\prime).
   \label{commutation-relation}
\eea
In order for Eq. (\ref{annihilation}) to be in accord with
Eq. (\ref{commutation-relation}), the mode function 
$\delta \phi_{\varphi {\bf k}} (t)$
should follow the Wronskian condition
\bea
   & & \delta \phi_{\varphi {\bf k}} \delta \dot \phi_{\varphi {\bf k}}^{*}
       - \delta \phi^*_{\varphi {\bf k}} \delta \dot \phi_{\varphi {\bf k}}
       = {i \over a^3 Z}.
   \label{Wronskian}
\eea

{\it Assuming}
\bea
   & & z^{\prime\prime} / z = n / \eta^2, \quad n = {\rm constant},
   \label{n}
\eea
Eq. (\ref{GGT-delta-phi-eq}) becomes a Bessel equation with a solution
\bea
   \delta \phi_{\varphi {\bf k}} (\eta) 
   &=& {\sqrt{ \pi |\eta|} \over 2 a} \Big[ c_1 ({\bf k}) H_\nu^{(1)} 
       (k|\eta|)
   \nonumber \\
   & & + \;
       c_2 ({\bf k}) H_\nu^{(2)} (k|\eta|) \Big] {1 \over \sqrt{Z}}, \quad
       \nu \equiv \sqrt{ n + {1\over 4} }.
   \label{delta-phi-k-sol}
\eea
The coefficients $c_1({\bf k})$ and $c_2({\bf k})$ are arbitrary functions 
of ${\bf k}$ which are normalized according to Eq. (\ref{Wronskian}) as
\bea
   & & | c_2 ({\bf k}) |^2 - | c_1 ({\bf k}) |^2 = 1.
   \label{normalization}
\eea
Imposition of the quantization condition in Eq. (\ref{Wronskian}) does not
completely fix the coefficient.
The remaining freedom is related to the choice of the vacuum state.
The adiabatic vacuum (in de Sitter space it is often called as 
Bunch-Davies vacuum, \cite{EXP}) chooses $c_2 ({\bf k}) \equiv 1$ and 
$c_1 ({\bf k}) \equiv 0$ which corresponds to the positive frequency 
solution in the Minkowski space limit.

The power spectrum becomes
\bea 
   {\cal P}_{\delta \hat \phi_\varphi} ( k, t ) 
   &\equiv& {k^3 \over 2 \pi^2} \int \langle \delta \hat \phi_\varphi
       ({\bf x} + {\bf r}, t) \delta \hat \phi_\varphi ({\bf x}, t)
       \rangle_{\rm vac} e^{- i {\bf k} \cdot {\bf r} } d^3 r
   \nonumber \\
   &=& {k^3 \over 2 \pi^2} | \delta \phi_{\varphi {\bf k}} (t) |^2,
   \label{Power-spectrum}
\eea
where we used $a_{\bf k} | {\rm vac} \rangle \equiv 0$ for every ${\bf k}$.
Assuming the adiabatic vacuum, the two-point function becomes 
\bea 
   & & G ( x^\prime, x^{\prime\prime} ) 
       \equiv \langle \delta \hat \phi_\varphi (x^\prime)
       \delta \hat \phi_\varphi (x^{\prime\prime}) \rangle_{\rm vac}
       = {({1\over 4} - \nu^2 ) \sec{(\pi \nu)} \over 16 \pi 
       a^\prime a^{\prime\prime} \eta^\prime \eta^{\prime\prime}}
   \nonumber \\
   & & \qquad \times
       F \left( {3\over2} + \nu, {3\over 2} - \nu; 2; 1 + {\Delta \eta^2 
       - \Delta {\bf x}^2 \over 4 \eta^\prime \eta^{\prime\prime} } \right)
       { 1 \over \sqrt{Z^\prime Z^{\prime\prime}} },
   \nonumber \\
   \label{Green-function}
\eea
which is valid for $\nu < {3\over 2}$; $x \equiv ({\bf x}, t)$, 
$\Delta \eta^2 \equiv (\eta^\prime - \eta^{\prime\prime})^2$ and
$\Delta {\bf x}^2 \equiv ({\bf x}^\prime - {\bf x}^{\prime\prime} )^2$.

In the small scale limit, thus $k \eta \gg 1$, (\ref{delta-phi-k-sol})
becomes
\bea
   \delta \phi_{\varphi {\bf k}} (\eta)
   &=& {1 \over a \sqrt{2k}}
       \Big[ c_1({\bf k}) e^{i k \eta - i ( \nu + {1\over 2} ) {\pi \over 2} }
   \nonumber \\
   & & \qquad \qquad + \;
       c_2 ({\bf k}) e^{- i k \eta + i ( \nu + {1\over 2} ) {\pi \over 2} }
       \Big] {1\over \sqrt{Z}}.
   \label{delta-phi-k-sol-SS}
\eea
In the large-scale limit we have
\bea
   & & \delta \phi_{\varphi {\bf k}} (\eta) 
       = i { \sqrt{|\eta|} \Gamma ( \nu ) \over a 2 \sqrt{\pi} } 
       \left( { k |\eta| \over 2} \right)^{-\nu} 
       \Big[ c_2 ({\bf k}) - c_1 ({\bf k}) \Big] {1\over \sqrt{Z}},
   \nonumber \\
   \label{delta-phi-k-sol-LS}
\eea
and the power spectrum becomes 
\bea 
   & & {\cal P}^{1/2}_{\delta \hat \phi_\varphi} ( k, \eta ) 
       = { \Gamma ( \nu ) \over \pi^{3/2} a |\eta| } 
       \left( { k |\eta| \over 2} \right)^{3/2-\nu} 
       \Big| c_2 ({\bf k}) - c_1 ({\bf k}) \Big| {1\over \sqrt{Z}}.
   \nonumber \\
   \label{Power-spectrum-LS} 
\eea 
In Eqs. (\ref{delta-phi-k-sol}-\ref{Power-spectrum-LS}) no additional
dependence on ${\bf k}$ arises from the generalized nature of the theory.

Let us see the implication of the condition in Eq. (\ref{n}). 
Introduce the following notations
\bea
   & & \epsilon_1 \equiv {\dot H \over H^2}, \quad
       \epsilon_2 \equiv {\ddot \phi \over H \dot \phi}, \quad
       \epsilon_3 \equiv {1 \over 2} {\dot F \over H F}, 
   \nonumber \\
   & & \epsilon_4 \equiv {1 \over 2} {\dot E \over H E}, \quad
       E \equiv F \left( \omega + {3 \dot F^2 \over 2 \dot \phi^2 F} \right).
\eea
{}For $\dot \epsilon_1 = 0$, we have 
\bea
   & & \eta = - {1\over aH} {1\over 1 + \epsilon_1},
   \label{eta-aH}
\eea
and for $\dot \epsilon_i = 0$ we have 
[for general $\epsilon_i$'s, see Eq. (88) of \cite{GGT-HN}]
\bea
   {z^{\prime\prime} \over z} = {n \over \eta^2}
   &=& {1\over \eta^2} {1\over ( 1 + \epsilon_1 )^2 }
       \Big( 1 - \epsilon_1 + \epsilon_2 - \epsilon_3 + \epsilon_4 \Big)
   \nonumber \\
   & & \quad \times
       \Big( 2 + \epsilon_2 - \epsilon_3 + \epsilon_4 \Big).
   \label{z-epsilon-const}
\eea
{}For $\dot \epsilon_i = 0$ we have $a \propto t^{-1/\epsilon_1}$,
$\dot \phi \propto a^{\epsilon_2}$ 
(thus, $\phi \propto t^{1-\epsilon_2/\epsilon_1}$ 
for $\epsilon_1 \neq \epsilon_2$, and
$\phi \propto \ln{t}$ for $\epsilon_1 = \epsilon_2$),
$F \propto a^{2 \epsilon_3}$, and $E \propto a^{2\epsilon_4}$.
{}For $a \propto t^q$ with $q = {2 \over 3 ( 1 + {\rm w})}$,
we have $\epsilon_1 = - 1/q$ and $a \propto \eta^{2/(1 + 3 {\rm w})}$.
In the limit of Einstein's gravity, thus $Z = 1$,
the solutions in Eqs. (\ref{delta-phi-k-sol}-\ref{Power-spectrum-LS}) 
reduce to the ones derived in \S III of \cite{H-QFT}, \cite{EXP,POW}.
{}For a power-law expansion stage supported by the scalar field in 
Einstein's gravity, we have $\epsilon_3 = 0 = \epsilon_4$ and
$\dot \phi / H = {\rm constant}$.
Thus, $\epsilon_1 = \epsilon_2 = - 1/q$ and 
Eqs. (\ref{delta-phi-k-sol},\ref{z-epsilon-const}) lead to
\bea
   & & \nu = { 3 q - 1 \over 2 ( q -1 ) } 
       = { 3 ({\rm w} -1) \over 2 ( 3 {\rm w} + 1 ) }.
\eea 
${\rm w} < - {1 \over 3}$ corresponds to $q>1$.
The exponential expansion stage corresponds to
${\rm w} \rightarrow -1$, thus $\nu \rightarrow {3 \over 2}$;
in this case we have $\eta = - 1 /(a H)$ where $H$ becomes a constant.

Now, we derive some observationally relevant classical power spectrums 
generated from quantum fluctuations as the initial seeds.
Ignoring the transient mode, from Eq. (\ref{UCG-delta-phi-LS-sol}) we have 
\bea
   & & {\cal P}^{1/2}_C (k,t) = {H \over |\dot \phi|}
       {\cal P}^{1/2}_{\delta \phi_\varphi} (k,t), 
   \label{Power-C}
\eea
where the classical power spectrum of a fluctuating field $f({\bf x}, t)$
is defined as
\bea
   & & {\cal P}_f ( k, t) \equiv
       {k^3 \over 2 \pi^2} \int \langle f ({\bf x} + {\bf r}, t)
       f ({\bf x}, t) \rangle_{\bf x} e^{- i {\bf k} \cdot {\bf r} } d^3 r.
   \label{P-spatial}
\eea
We have in mind a generic scenario where the classical structures arise 
from the quantum fluctuations pushed outside horizon and classicalized 
during accelerated expansion stage.
As an {\it ansatz} we identify 
\bea
   & & {\cal P}_{\delta \phi_\varphi} (k,t) 
       = Q(k,t) \times {\cal P}_{\delta \hat \phi_\varphi} (k,t),
   \label{ansatz}
\eea
where ${\cal P}_{\delta \phi}$ and ${\cal P}_{\delta \hat \phi}$ 
are based on the classical volume average 
and the quantum vacuum expectation value, respectively,
Eqs. (\ref{P-spatial},\ref{Power-spectrum}). 
$Q(k,t)$ is a factor which may take into account of the possible 
modification of the spectrum due to the 
classicalization process of the quantum field fluctuations.
We may call it a ``classicalization factor''.
Ordinarily it is taken to be unity, however, the decoherence,
noise and nonlinear field effects may affect its value, 
particularly its amplitude, \cite{Hu}.
Assuming Eq. (\ref{n}) we have derived the quantum fluctuations in 
the large scale limit in Eq. (\ref{Power-spectrum-LS}).
Thus, combining Eqs. (\ref{Power-C},\ref{ansatz},\ref{Power-spectrum-LS}) 
we have
\bea
   {\cal P}^{1/2}_C (k, \eta) 
   &=& {H \over |\dot \phi|}
       { \Gamma ( \nu ) \over \pi^{3/2} a |\eta| }
       \left( { k |\eta| \over 2} \right)^{3/2-\nu}
   \nonumber \\
   & & \times \Big| c_2 ({\bf k}) - c_1 ({\bf k}) \Big| 
       \sqrt{Q (k)\over Z (\eta)} \Bigg|_{\rm LS,GGT},
   \label{Power-C-QF}
\eea
where quantities in the right hand side should be evaluated when 
the scale we are considering was in the large scale limit (LS) during 
an expansion stage supported by a generalized gravity theory (GGT).
In the Einstein gravity limit ($Z =1$) and an exponential expansion (EXP) 
stage [$\nu = {3 \over 2}$ and $\eta = -1/(aH)$] Eq. (\ref{Power-C-QF}) 
reduces to the well known result
\bea
   & & {\cal P}^{1/2}_C (k, \eta) = {1 \over 2 \pi} {H^2 \over |\dot \phi|}
       \Big| c_2 ({\bf k}) - c_1 ({\bf k}) \Big| \sqrt{Q (k)} 
       \Bigg|_{\rm LS,EXP}.
   \label{Power-C-QF-EXP}
\eea
We note that in general the power spectrum depends on the choice of a
vacuum state and possibly on the classicalization factor $Q (k)$.

Now, in Eq. (\ref{Power-C-QF}) we have the ``quantum fluctuation generated 
power spectrum'' imprinted in a conserved quantity $C({\bf x})$.
{}From the power spectrum of $C({\bf x})$ we can derive the spectrums of
observable quantities in the present universe, e.g., density fluctuations,
velocity fluctuations, potential fluctuations, and temperature 
fluctuations in the cosmic microwave background radiation
in the matter dominated era in Einstein's gravity;
these are \cite{GGT-H}
\bea
   & & {\delta \varrho \over \varrho} 
       = {2 \over 5} \left( {k \over a H} \right)^2 C, \quad
       \delta v = - {2 \over 5} \left( {k \over aH} \right) C, 
   \nonumber \\
   & & 
       \delta \Phi = - {3 \over 5} C, \quad
       {\delta T \over T} = {1\over 5} C.
   \label{obs-fluctuations}
\eea
Notice that we are considering a linear theory.
In the linear theory, all perturbed order quantities are linearly 
related with each other which is true even between variables in 
different gauges.
$C({\bf x})$ is a temporally constant, but spatially varying, 
coefficient of the growing mode.
The spatial curvature (or potential) fluctuation in the uniform-field
gauge (which coincides with the comoving gauge [CG] in a minimally coupled
scalar field, see Sec. IV C of \cite{GGT-HN}) is conserved as $C({\bf x})$;
i.e., $\varphi_{\delta \phi} ({\bf x}, t) = C({\bf x}) 
= \varphi_{\rm CG} ({\bf x}, t)$.
$C({\bf x})$ encodes the spatial structure of the fluctuations and 
is conserved during the linear evolution in the large scale limit.
Indeed, this linearity is one basic underlying reason why we were 
successful in tracing the structure evolution in a simple and unified way.
However, apparently, there arises no structure formation in the linear 
theory [structures are preserved in $C({\bf x})$],
and one should not miss that the gravity theories are highly nonlinear.

In this paper we have not used the conformal transformation which relates 
the generalized gravity in Eq. (\ref{Action-GGT}) to Einstein's gravity 
in classical level, \cite{GGT-CT}.  
Quantum fluctuations are derived in the original frame of the generalized 
gravity. 
Still, the underlying conformal symmetry with Einstein's gravity can be 
regarded as an important factor which allows the unified and simple 
analyses possible in the classical level, \cite{GGT-CT}.

We have implictly assumed the existence of an accelaration stage 
supported by the generalized gravity with the condition in Eq. (\ref{n}). 
Constructing specific models with applications will be considered elsewhere.

We thank Dr. H. Noh for useful discussions.
This work was supported in part by the Korea Science and Engineering
Foundation, Grants No. 95-0702-04-01-3 and No. 961-0203-013-1, 
and through the SRC program of SNU-CTP.


\end{document}